# Mechanical properties of AlMgB$_{14}$-related boron carbide structures. A first principle study

Oleksiy Bystrenko [a,1], Jingxian Zhang [a,b], Dong Fangdong [c], Xiaoguang Li [a,b], Weiyu Tang [a,b], Kaiqing Zhang [a,b], Jianjun Liu [a,b]

[a] Key Laboratory of Advanced Structural Ceramics and Ceramics Matrix Composites, Shanghai Institute of Ceramics, Chinese Academy of Sciences, Shanghai 200050, China

[b] Center of Materials Science and Optoelectronics Engineering, University of Chinese Academy of Sciences, Beijing 100049, China

[c] Science and Technology on Transient Impact Laboratory, Beijing, 102202, China

Abstract:    We examine the effects produced by replacing B-B interlayer bonds by C-C bonds in AlMgB$_{14}$-related boron network on its mechanical properties. The elastic constants, Vickers hardness and shear strength are evaluated by means of first principle computer simulations on the basis of density functional theory. The results of simulations suggest a possibility of existence of several orthorhombic boron carbide phases with strongly enhanced mechanical properties with the Young's modulus and Vickers hardness being within the range of 550- 600 GPa and 43-50 GPa, respectively.

## I. Introduction

In recent years, since the publication of Ref. [1], the properties of orthorhombic metal borides of the composition XYB$_{14}$ (where X, Y are metal atoms) attracted considerable attention of researchers. As was reported in Ref. [1], the aluminum magnesium boride AlMgB$_{14}$ (BAM) has the remarkable Vickers hardness of nearly 30 GPa, which, with addition of Si or TiB$_2$, can be increased up to 35 or 46 GPa, respectively. It is assumed that this significant hardness increase could be associated with the effects of microstructural nano-composite hardening or with the effect of doping by impurities present in specimens due to the mechanical alloying method of their preparation.

The above cited experimental results initiated a number of theoretical works, where the mechanical and electronic properties of BAM-based structures, in particular, modified by replacing elements on metal sites were examined on the basis of first principle computer simulations [2-5]. In Ref.[3], to understand the mechanisms of high hardness of BAM-related compounds, the calculations of cleavage strength in different orientations for AlLiB$_{14}$ were performed, and in Refs. [4, 5] the shear and tensile strength in different directions was examined.

The structure of BAM consists of boron layers formed by boron icosahedra B$_{12}$ connected by strong interlayer B-B bonds (which directly connect icosahedra of different layers), inter-icosahedra boron atoms connecting the icosahedra within a layer, and metal atoms embedded on metal sites X and Y, as shown in Fig.1 [6]. In

---

contrast to most known superhard materials, BAM has low symmetry (Imma [74]), and relatively low density, which raises the question about the origin of its high hardness.

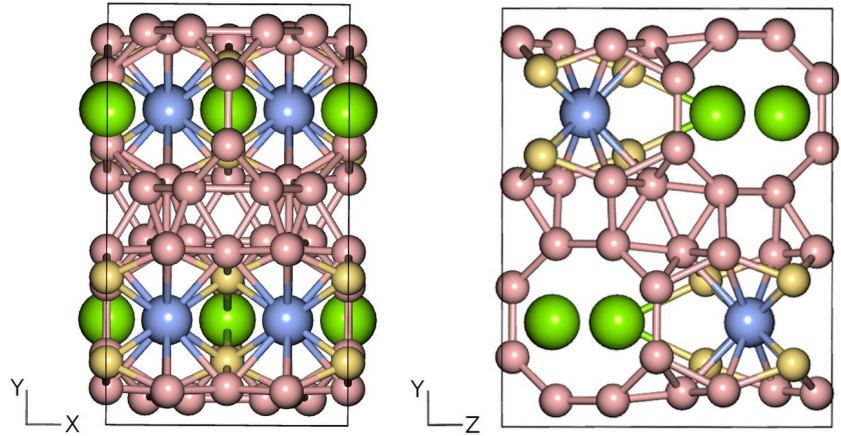

Fig. 1. Structure of AlMgB$_{14}$. Magnesium (X-sites), aluminum (Y-sites) and icosahedra boron atoms are given in green, blue and light brown, respectively. Inter-icosahedra boron is given in yellow.

The important conclusion of Ref. [3] is that, due to the low interlayer bond density, the weakest cleavage planes in BAM-related structure are exactly those parallel to the boron layers, which pass between them and cross these interlayer bonds. From this, the authors conclude that the enhancement of mechanical properties of BAM-related materials can be achieved by introducing dopants, which directly affect these bonds.

In this context, it is of interest to consider the possible effects of replacing boron atoms forming interlayer B-B bonds by carbon on the mechanical properties and structure of BAM-related compounds in more detail. We examine this issue by means of first principle computer simulations. As follows from the results of calculations, the boron carbide structures obtained in such a way can manifest very significant enhancement of mechanical properties, in particular, increase in theoretically predicted Young's modulus and hardness.

**II. Computer simulations.**

In simulations, we considered the following BAM-related systems: 1) aluminum magnesium boride AlMgB$_{14}$, as the basic reference system, 2) pure boron phase B$_{56}$, which represents BAM structure with vacancies on all metal sites; 3) structure B$_{40}$C$_{16}$, which differs from B$_{56}$ in that the 16 boron atoms forming the bonds between the boron layers (including 8 boron atoms on polar sites of icosahedra and 8 inter-icosahedra boron atoms) are replaced by carbon; 4) structure B$_{40}$C$_{20}$, which can be obtained from the previous structure B$_{40}$C$_{16}$ by adding 4 carbon atoms on empty metal Y sites; 5) structure B$_{44}$C$_{16}$, which can be obtained from the structure B$_{40}$C$_{16}$ by adding 4 boron atoms on empty metal Y sites. First principle simulations were carried out on the basis of density functional theory (DFT) by using free Quantum Espresso software [7]. In calculations, the standard solid state pseudo-potential library (SSSP: precision, version 1.1.2 [8]) with the exchange-correlation energy given in Perdew–Burke-Ernzerhof form [9] was employed. The uniform Monkhorst-Pack grid [10]

6x4x4 (where the first number relates to the direction of the shortest lattice dimension) with no shift was used. Selected runs were carried out on 5x3x3 and 4x2x2 k-point grids to control the accuracy of simulations. The cutoff energy was 55 Ry for wave functions, and 440 Ry for charge; the total energy convergence threshold for the electron wave functions was set $10^{-6}$ Ry.

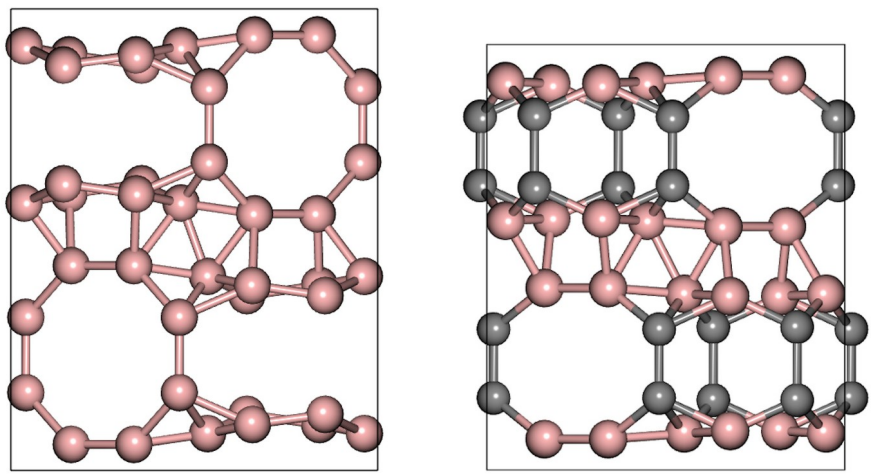

Fig. 2. Optimized structures of pure boron phase $B_{56}$ (left) and similar boron carbide structure $B_{40}C_{16}$ with all interlayer B-B bonds replaced by C-C bonds (right).

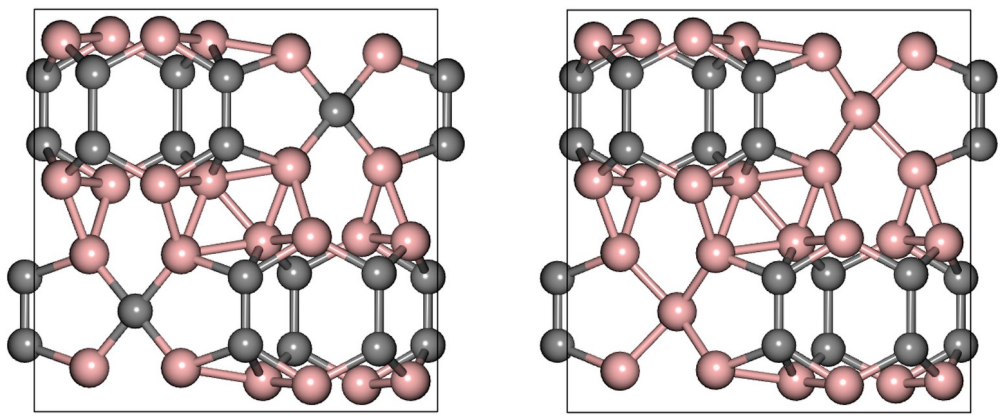

Fig. 3. Optimized boron carbide structures $B_{40}C_{20}$ (left) and $B_{44}C_{16}$ (right).

Initial information for BAM-related structures was taken from the materials database [11] with the orthorhombic unit cell represented by 64 atoms. The initial configurations for the structures to be considered were prepared by removing, adding, or replacing selected atoms on specified sites. In all cases, the simulations were preceded by optimization (relaxation) of structures with respect to ion positions and all lattice parameters for free cell geometry (i.e., including all cell dimensions and angles). Optimization was performed by means of BFGS method [12] and was aimed at finding the equilibrium ion configuration and lattice geometry at zero temperature and zero pressure associated with minimum energy. To determine the symmetry of calculated optimized structures, the free software package FINDSYM [13] with the ionic position tolerance of 0.002 angstrom was used. The final optimized structures 1) - 5) are displayed in Figs. 1 – 3, and the calculated structural data are given in Table I.

| No | Composition, number of atoms, $N_{at}$ | Supercell crystal type, symmetry, space group number | Supercell parameters, angstrom | Supercell volume; (vol. per atom), ang$^3$ | Atom | Wyckoff notation | Fractional coordinates | | |
|---|---|---|---|---|---|---|---|---|---|
| | | | | | | | x | y | z |
| 1 | AlMgB$_{14}$ $N_{at}=64$ | Orthorhombic Imma [74] | a=5.907 b=10.346 c=8.110 | 495.66 (7.74) | Mg Al B(1) B(2) B(3) B(4) B(5) | 4e 4d 16j 16j 8h 8h 8h | 0.000 0.250 0.160 0.746 0.000 0.000 0.000 | 0.250 0.250 0.063 0.581 0.087 0.149 0.167 | 0.152 0.750 0.336 0.046 0.672 0.883 0.469 |
| 2 | B$_{56}$ $N_{at}=56$ | Orthorhombic Imma [74] | a=5.618 b=10.401 c=8.332 | 486.93 (8.70) | B(1) B(2) B(3) B(4) B(5) | 16j 16j 8h 8h 8h | 0.163 0.748 0.000 0.000 0.000 | 0.556 0.080 0.599 0.621 0.668 | 0.335 0.036 0.659 0.857 0.459 |
| 3 | B$_{40}$C$_{16}$ $N_{at}=56$ | Orthorhombic Imma [74] | a=5.494 b=9.290 c=7.874 | 401.90 (7.18) | B(1) B(2) B(3) C(1) C(2) | 16j 16j 8h 8h 8h | 0.163 0.765 0.000 0.000 0.000 | 0.072 0.586 0.092 0.164 0.169 | 0.339 0.052 0.680 0.867 0.481 |
| 4 | B$_{40}$C$_{20}$ $N_{at}=60$ | Orthorhombic Imma [74] | a=5.638 b=8.526 c=8.492 | 408.21 (6.80) | C(1) C(2) C(3) B(1) B(2) B(3) | 4d 8h 8h 16j 16j 8h | 0.250 0.000 0.000 0.663 0.270 0.000 | 0.250 0.156 0.165 0.609 0.076 0.066 | 0.750 0.355 0.024 0.364 0.067 0.188 |
| 5 | B$_{44}$C$_{16}$ $N_{at}=60$ | Orthorhombic Pmma [51] | a=8.679 b=5.520 c=8.653 | 414.57 (6.91) | B(1) B(2) B(3) B(4) B(5) B(6) B(7) C(1) C(2) C(3) C(4) | 4k 8l 8l 8l 8l 4j 4i 4j 4i 4j 4i | 0.250 0.105 0.897 0.078 -0.078 0.067 -0.071 0.156 0.844 0.166 0.834 | 0.169 0.336 0.161 0.233 0.270 0.500 0.000 0.500 0.000 0.500 0.000 | 0.773 0.868 0.368 0.068 0.566 0.187 0.685 0.353 0.852 0.025 0.524 |

Table I. Structural properties of AlMgB$_{14}$ and related boron carbide phases, obtained in simulations.

The optimized configurations were used then to evaluate cohesive and formation energy, elastic properties and hardness of the above BAM-related phases. Formation energy was calculated with respect to pure elemental phases of carbon as α-graphite (P6$_3$/mmc), boron as α-B$_{12}$ (R-3m), hexagonal magnesium (P6$_3$/mmc), and cubic aluminum (Fm-3m). Elastic moduli B, G, Young's modulus E and Poisson's ratio

$\nu$ were determined with the use of free THERMO_PW software [14], which employs strain-stress approach for a set of different strains to evaluate elastic constants.

Let us say a few words about evaluating hardness on the basis of computer simulations. Hardness is a macroscopically defined quantity, and, therefore, depends on a large number of conditions like the type of indentation process, specimen microstructure, presence of defects, etc. For this reason, all theoretical approaches to find hardness including those based of microscopic first principle calculations proposed so far [15-20] remain semi-empirical and can give only approximate numbers. We used for this purpose the method of Ref. [20] based on determining the elastic constants and Poisson's ratio, which was shown to provide reasonable agreement with known experimental data for a large number of superhard materials.

The results of calculations are given in Table II. In all cases the resulting optimized lattice geometry remained orthorhombic while the atomic configurations in the vicinity of metal sites Y for the structures 4) and 5) were significantly modified in the process of relaxation.

The most remarkable finding is that due to replacement of B-B interlayer bonds by C-C bonds, the BAM-related structures undergo a significant contraction along the direction perpendicular to boron layers, accompanied by essential increase in elastic moduli. Notice that the numbers 550-600 GPa for Young's modulus are typical for ultrahard materials. The theoretical estimates of Vickers hardness for these structures give the numbers within the range of 43-50 GPa.

| No | Composition | Density g/cm$^3$ | Cohesive energy per atom eV | Formation energy per atom eV | Elastic moduli, GPa | | | Poisson's ratio, $\nu$ | Vickers hardness GPa |
|---|---|---|---|---|---|---|---|---|---|
| | | | | | Bulk modulus B | Shear modulus G | Young's modulus, E | | |
| 1 | AlMgB$_{14}$ | 2.72 | -6.151 | -0.089 | 201.2 | 193.7 | 440.0 | 0.14 | 33.4 |
| 2 | B$_{56}$ | 2.06 | -6.336 | 0.230 | 130.9 | 65.3 | 167.6 | 0.28 | 8.4 |
| 3 | B$_{40}$C$_{16}$ | 2.58 | -7.260 | 0.007 | 249.9 | 244.5 | 553.0 | 0.13 | 43.5 |
| 4 | B$_{40}$C$_{20}$ | 2.74 | -7.209 | 0.174 | 274.4 | 275.7 | 619.6 | 0.12 | 50.1 |
| 5 | B$_{44}$C$_{16}$ | 2.68 | -7.135 | 0.085 | 255.1 | 266.0 | 592.3 | 0.11 | 49.0 |

Table II. Cohesive and formation energy and mechanical properties of BAM-related structures, calculated on the basis of DFT simulations. For the elastic moduli the Voigt-Reuss-Hill averages are given.

The examination of electronic properties of the structures under consideration has been done in a usual way with parameters given above. The comparison of the density of states (DOS) is given in Fig. 4.

According to the results presented in the figure, one should expect the boron carbide structures B$_{40}$C$_{16}$, B$_{40}$C$_{20}$, B$_{44}$C$_{16}$ to be semiconductors or insulators with band gap ranging from nearly zero for B$_{40}$C$_{20}$ to appoximately 2 eV for B$_{40}$C$_{16}$ (notice that DFT simulations systematically underestimate the band gap, thus, these numbers are approximate).

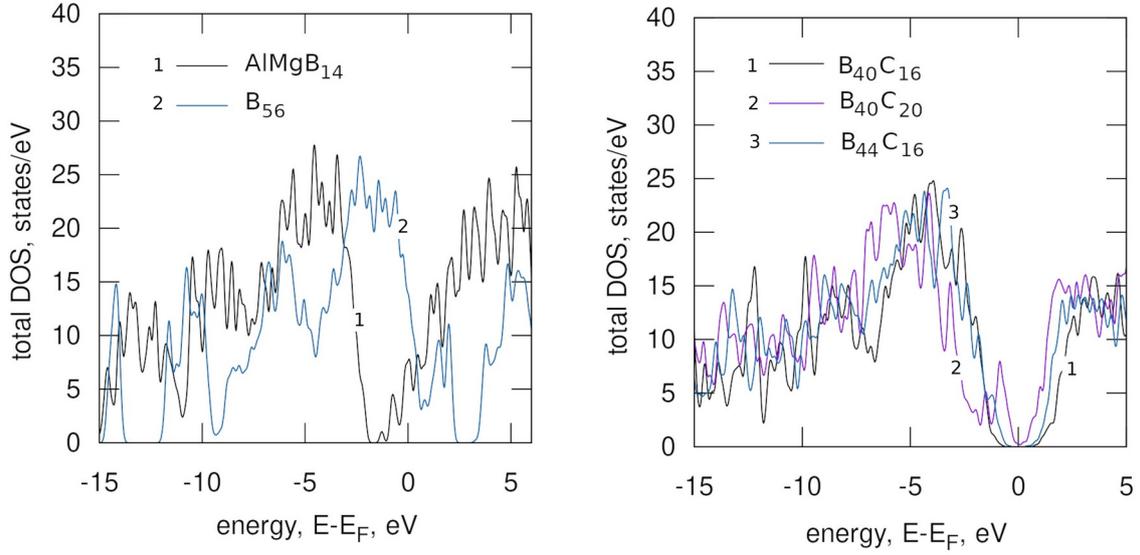

Fig. 4. Comparison of total density of states for $AlMgB_{14}$ and boron phase $B_{56}$ (left) and orthorhombic boron carbide structures $B_{40}C_{16}$, $B_{40}C_{20}$, $B_{44}C_{16}$ (right).

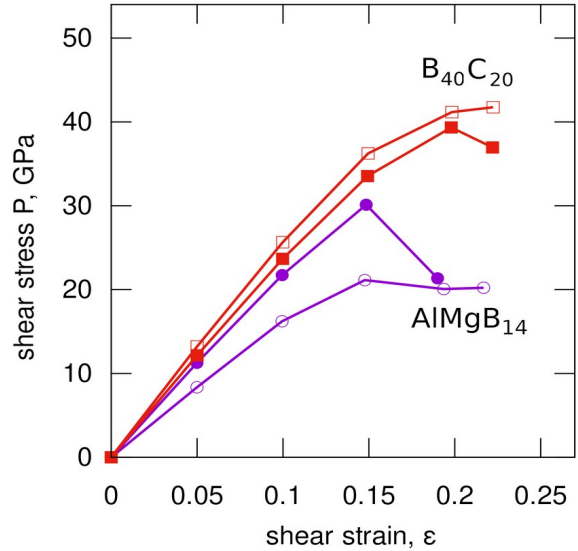

Fig. 5. Shear stress-strain dependencies for $AlMgB_{14}$ (blue circles) and boron carbide structure $B_{40}C_{20}$ (red squares). Open and filled symbols relate to the weakest (010)[100] and strongest (001)[100] shear systems, respectively.

It is commonly accepted that hardness is closely related to the shear strength of materials. For this reason, it is of interest to examine this quantity for the above structures in direct computer simulations. According to the results of Refs [3, 5], the weakest shear system in conventional BAM-related materials associated with relative sliding of boron layers should be expected as (010)[100], and the strongest shear system should be (001)[100] (with the coordinate system defined as in Fig. 1). As an example, we examined the process of elastic failure under shear strain in $B_{40}C_{20}$ and $AlMgB_{14}$ (as the reference system) in the following way. The unit cell was subjected to a series of subsequently increasing uniform shear deformations within a selected shear system. At each step, ionic positions and supercell dimensions were relaxed in such a way as to keep the selected shear system fixed. The relaxation of supercell dimensions is needed to obtain the strain-stress dependencies at fixed pressure (i.e., to keep the overall pressure close to zero). After structure relaxation, at each step, the stress tensor was evaluated and strain re-calculated.

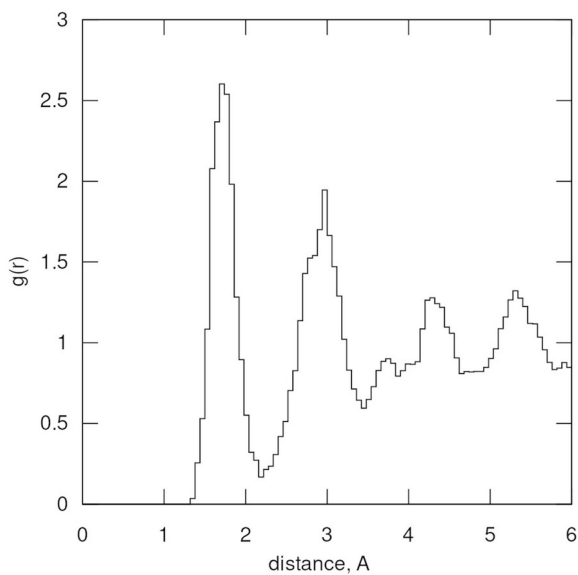

Fig. 6. Radial distribution function for boron carbide structure $B_{40}C_{20}$ calculated for the temperature T= 1200 C by quantum molecular dynamics.

As follows from the result of simulations (Fig. 5), the boron carbide structure $B_{40}C_{20}$ demonstrates much higher shear strength than the reference system $AlMgB_{14}$ in both weakest and strongest shear directions (nearly 40 GPa in both directions for $B_{40}C_{20}$ vs. 20 and 30 GPa for $AlMgB_{14}$). Thus, these results support the theoretical predictions for high hardness of boron carbide $B_{40}C_{20}$ and correlate with the conclusions of Ref. [3].

The formation energies for the hypothetical superhard boron carbide structures 3)-5) are positive, which means that these phases are meta-stable at zero pressure and temperature. In this context it is of interest to examine the thermal stability of the above compounds at non-zero temperature. It was done by means of quantum Born-Oppenheimer molecular dynamics simulations. The computations were carried out for the boron carbide phases for the temperature T=1200 C and zero pressure for free cell geometry, i.e., the dynamics of cell dimensions and angles was taken into account as well. The result of calculation of binary radial distribution function for the boron carbide phase $B_{40}C_{20}$ is given in Fig. 6. The total time span of simulation was 0.5 ps and the averaging was performed over the 200 final configurations during the time span 0.4-0.5 ps. The results of simulations indicate that, after some transition period of 0.06 ps associated with thermalization of the system, it remains then in steady state at this temperature. At this, no any phase change occurs, and the system retains its initial crystal structure (which is also evident from the direct visual observations of configurations in the process of molecular dynamics simulations). Similar behavior was established for the structures 3) and 5). To find the conditions for the thermodynamic stability of the hypothetical boron carbide phases considered, a study of phase diagram is needed, which, however, is the subject for an independent research.

**Conclusions**

To conclude, we theoretically examined the effects produced by replacing interlayer B-B bonds in AlMgB$_{14}$-related boron network by C-C bonds by means of first principle DFT simulations. The obtained results suggest a possibility of existence of a number of orthorhombic boron carbide phases with non-trivial mechanical properties with the theoretically predicted Young's modulus of 550-600 GPa and intrinsic hardness 43-50 GPa, which may provide the basis for developing novel superhard materials.

The obtained results support the conclusions of Ref. [3] concerning the importance of interlayer B-B bonds for mechanical properties of BAM-related materials.

**Acknowledgment**

The authors gratefully acknowledge the support of this research within the framework of the CAS President's International Fellowship Initiative, grant No 2020VEB0005.